\documentclass[11pt,twoside,twocolumn,english]{article}
\usepackage{mathptmx}
\usepackage[T1]{fontenc}
\usepackage[utf8]{inputenc}
\usepackage[a4paper]{geometry}
\usepackage{color}
\usepackage{babel}
\usepackage{mathrsfs}
\usepackage{amsmath}
\usepackage{amssymb}
\usepackage{esint}
\usepackage[unicode=true,pdfusetitle,
 bookmarks=true,bookmarksnumbered=false,bookmarksopen=false,
 breaklinks=false,pdfborder={0 0 1},backref=false,colorlinks=true]
 {hyperref}
\usepackage{breakurl}

\makeatletter

\newcommand*\LyXThinSpace{\,\hspace{0pt}}

\@ifundefined{date}{}{\date{}}
\usepackage{babel}
\usepackage{babel}
\usepackage{babel}
\allowdisplaybreaks[4]

\makeatother

\begin{document}
\global\long\def\divg{{\rm div}\,}

\global\long\def\curl{{\rm curl}\,}

\global\long\def\rt{\mathbb{R}^{3}}

\global\long\def\rn{\mathbb{R}^{n}}

\global\long\def\dir{\mathcal{E}}

\title{Spacetime orientation and the meaning of\\
 Lorentz invariance in general relativity}

\author{Zhongmin Qian\thanks{Exeter College, Turl Street, Oxford, OX1 3DP, England. Email: $\mathtt{zhongmin.qian@exeter.ox.ac.uk}$}}
\maketitle

\noindent{\textbf{Summary}}. The parity violation at the level of weak interactions and other similar
discrete symmetries breaking show that the invariance of laws under
the full group of Lorentz transformations can not be taken granted.
We examine the principle of Lorentz invariance under the general theory
of relativity, and demonstrate the importance of a concept of the
spacetime orientation as part of the causal structure of spacetime. 

\section{Introduction}

According to French 
(see page 66, \cite{French}), ``a relativity principle is an assertion
about the laws of nature as they would be determined by observations
made in different frames of reference'', and Galileo and Newton's
relativity principle is ``the assertion ... that there are whole classes
of reference frames with respect to which the laws of physics have
precisely the same form''. In the review article \cite{Bondi 1959},
H. Bondi stated that ``A physical statement of what these invariants
are is called a principle of relativity, and the fundamental equations
of a theory usually define the principle of relativity applicable
to it.'' Therefore Bondi recognized the content of  invariance 
principles to be included in the principle of relativity.

Throughout the history, the principle of relativity has gone through
several critical revisions by the scientific giants, and each crucial
revision over the last 300 years, from Galileo, Newton to Einstein,
has brought a revolution in mankind's understanding of laws of Nature.
According to the current research in science and technology, it is
recognized that the principle of relativity, formulated by Einstein
in the most recent revision of the principle, covers three postulates
which have different physical meanings. The first is The Principle
of Equivalence, which leads to the mathematical models of space-time.
The second is the so-called The Covariance Principle, which determines
possible mathematical formulation for laws of Nature, but does not
involve physical processes. The last one, without doubt the most
important component of the principle of relativity, is The Principle
of Lorentz Invariance, whose formulation itself, will be examined
carefully in the setting of general relativity in this article,
depends heavily on the gravitational potential, and is an important guide
for determining physical laws of Nature.

We shall begin with what Einstein said in his foundation papers on
the relativity theory. In the special relativity paper \cite{Einstein-special}, published in 1905 (see e.g. page 41, \cite{Lorentz}), Einstein formulated two
postulates to advance his theory of relativity.

1. The laws by which the states of physical systems undergo change
are not affected, whether these changes of state be referred to the
one or the other of two systems of coordinates in uniform translatory
motion.

2. Any ray of light moves in the ``stationary'' system of co-ordinates
with the determined velocity $c$, whether the ray be emitted by a
stationary or by a moving body.

The first postulate called the principle of relativity in Einstein
\cite{Einstein-special}, is more often called the principle
of equivalence, satisfied also by the mechanics of Galileo and Newton.
According to Eddington \cite{Eddington Math Theory}, the principle
of equivalence demands ``the laws of motion of undisturbed material
particles and of light-pulses in a form independent of the coordinates
chosen.'' (page 39, \cite{Eddington Math Theory}). It was already
clear to Eddington that the principle of equivalence ``has played
a great part as a guide in the original building up of the ... relativity
theory; but now ... it has become less necessary.'' (see page 41,
\cite{Eddington Math Theory}). Therefore, with the establishment
of mathematical models for spacetime, the principle of equivalence
is no longer useful for describing physics laws. In his foundation
paper \cite{Einstein-general} on general relativity, Einstein re-examined
the principle of relativity, and he wrote ``The laws of physics must
be of such a nature that they apply to systems of reference in any
kind of motion.'' Note here Einstein used the phase ``systems
of reference'' instead of ``coordinate systems''. In the same paper
(see page 117, \cite{Lorentz}), however, Einstein stated that ``The
general laws of nature are to be expressed by equations which hold
good for all systems of co-ordinates, that is, are co-variant with
respect to any substitutions whatever (generally co-variant).'' This
version of principle of relativity is too narrow which deprived the
important component in the previous formulation Einstein stated. It
is not known why Einstein retreated from the principle of relativity,
 to a narrow version
of the covariance principle.

In the only relativity textbook \cite{Bergmann} endorsed by Einstein,
Bergmann wrote (see page 154) that ``The principle of equivalence
was ostensibly a fundamental property of the gravitational forces'',
and ``the equivalence of gravitational and interial fields (which
is a consequence of the equality of gravitational and inertial masses)
gave the principle of equivalence its name''. About the principle
of general covariance, Bergmann stated (on page 159, \cite{Bergmann})
that ``The hypothesis is that the geometry of physical space is represented
best by a formalism which is covariant with respect to general coordinate
transformations, and that a restriction to a less general group of
transformations would not simplify that formalism, is called the 
principle
of general covariance''. Therefore, the equivalence principle is used to construct mathematical models for spacetime only, 
while the principle of covariance is a mathematical requirement for
constructing fields.

Let us look at what Pauli said about the principle of equivalence.
On page 145, \cite{Pauli}, Pauli formulated the general version of
the equivalence principle as that ``For every infinitely small world
region (i.e. a world region which is so small that the space- and
time-variation of gravity can be neglected in it) there always exists
a coordinate system – in which gravitation has no influence either
on the motion of particles or any other physical processes.'' What
Pauli stated, wrote just after the creation of Einstein's general
theory of relativity, is a very precise statement of the equivalence
principle – the gravitational potential can not be detected locally.

In \cite{weyl-spacetime matter}, Weyl formulated the principle of
relativity of Galilei and Newton as the following (page 154, \cite{weyl-spacetime matter}):
``it is impossible to single out from the systems of reference that
are equivalent for mechanics and of which each two are correlated
by the formulae of transformation III special systems without specifying
individual objects'', here a transformation III means a linear transformation
of following type 
\begin{align*}
x_{1} & =a_{11}x'_{1}+a_{12}x'_{2}+\gamma_{1}t'+\alpha_{1}\\
x_{2} & =a_{21}x'_{1}+a_{22}x'_{2}+\gamma_{2}t'+\alpha_{2}\\
t & =t'+a
\end{align*}
where $a_{ij}$, $\gamma_{k}$, $\alpha_{k}$ and $a$ are constant.
Thus Weyl included the principle of Lorentz invariance (though in
a restricted sense) as a component of relativity principle.

\section{The principle of relativity}

We are
in a position to give precise meanings for the context of the principle
of relativity, to motivate our interpretation of Lorentz invariance in
the general theory of relativity. 

\vskip0.3truecm

\emph{1. The equivalence principle and the principle of constancy
of light velocity.} The mathematical models of spacetime in special
and general theories of relativity (see \cite{Einstein-special,Einstein-general})
were established based on two postulates, the first is the equivalence
principle which does not involve any physics (see e.g. Norton
\cite{Norton} for a review about the meaning of this postulate), the second is the principle of constancy of light velocity,
which implicitly requires that physical processes propagate with speed
less than or equal to the velocity $c$ of light. Together, in the
special theory of relativity, the two postulates lead to the mathematical
model of spacetime, the Minkowski space. If the motion of a particle
is described by space coordinates $x,y$ and $z$, parameterized by
a time variable $t$, then the postulate of constancy of light speed
demands that $\dot{x}^{2}+\dot{y}^{2}+\dot{z}^{2}\leq c^{2}$, which
can be written in terms of differentials, i.e. $c^{2}dt^{2}-dx^{2}-dy^{2}-dz^{2}\geq0$.
The last inequality immediately leads to the Minkowski metric on $\mathbb{R}^{4}$,
so concepts of space and time are united in a natural way in this
model. In the general theory of relativity, the two postulates are
applied to develop general continuum space-time models being manifolds
of dimension four, together with a Lorentz metric $ds^{2}=g_{\mu\nu}dx^{\mu}dx^{\nu}$
with signature $(1,-1,-1,-1)$. The core element of the general theory
of relativity is however the postulate that the gravitation is described
by the curvature tensor field associated with the gravitational potential
$(g_{\mu\nu})$.

Taking a Lorentz manifold $(M,g_{\mu\nu})$ as a model of spacetime,
the principle of equivalence is no longer a postulate rather than
a mathematical statement: for every event  $x\in M$, there is
a local coordinate system about $x$ (i.e. there is a system of reference
near $x$), such that the gravitational potential $(g_{\mu\nu})$
is diagonalized to coincide with the Minkowski metric $\textrm{diag}(1,-1,-1,-1)$
(but this may be true only at $x$). Therefore, the gravitational
potential field can not be detected locally. The gravitational effect however
can not be wiped away, this is because for whatever the local coordinate
system we choose to describe physical processes, one can not write
off the curvature even locally.

\vskip0.3truecm

\emph{2. The principle of covariance}. Let us examine the
covariance principle for Einstein's spacetime $(M,g_{\mu\nu})$, which
is a principle emphasized in literature, it is however a basic requirement
for describing physical processes in a way mathematically meaningful.
Upon acceptance of the model for event space as a four dimensional
manifold $M$, even without specifying the gravitational potential
$(g_{\mu\nu})$, the principle of covariance, under the
current setting of theoretical physics, demands that fields describing
physical processes, either classical or quantized, must be sections
of fiber bundles over the four dimensional manifold $M$ of spacetime.
If derivatives of fields (to formulate dynamic equations of fields)
are required, then only covariant derivatives (including exterior
derivatives) are allowed, so that additional source of fields of differential
one forms, i.e. vector Boson fields,  must be introduced in to the physical processes. Therefore,
the covariance principle itself can not be applied to determine physics
laws.

\vskip0.3truecm

\emph{3. The principle of Lorentz invariance}. The important component
of the relativity principle is the principle of Lorentz invariance
for laws of Nature. In fact, the geometric dynamics, including Einstein's
field equation, has been developed under the help of \emph{a strong
version} of the Lorentz invariance in both special and general theories
of relativity, which can be stated that geometrical dynamic equations
describing\emph{ macroscopic motions} are invariant under diffeomorphisms
which preserve the gravitational field $(g_{\mu\nu})$. Einstein showed
this strong form of Lorentz invariance of the Maxwell equations, with
which the geometric dynamics were developed later by Planck and others.

Let us examine the principle of Lorentz invariance for the special
theory of relativity. Recall that the isometry group of the Minkowski
spacetime can be identified with the Poincar{é} group $\mathscr{P}$.
A homorphism $\varLambda$ of the Minkowski spacetime is an isometry
if it also preserves the Minkowski metric $(\eta_{\mu\nu})=(1,-1,-1,-1)$.
Under the standard coordinate system $(x^{\mu})$ in $\mathbb{R}^{4}$,
an isometry must take the form $\left(\varLambda x\right)^{\mu}=\varLambda_{\;\nu}^{\mu}x^{\nu}+a^{\mu}$,
where $(\varLambda_{\;\nu}^{\mu})$ is a constant matrix such that
$\varLambda_{\;\mu}^{\sigma}\eta_{\sigma\rho}\varLambda_{\;\nu}^{\rho}=\eta_{\mu\nu}$
(such a transformation is called a Lorentz transformation) and $a=\left(a^{\mu}\right)$
is a constant vector in $\mathbb{R}^{4}$. Therefore, the isometry
group in the special relativity is identified with the Poincar{é}
group $\mathscr{P}=\mathscr{L}\oplus\mathbb{R}^{4}$, where $\mathscr{L}=\textrm{SO}(1,3)$
is the Lorentz group. For the geometric dynamics, it has been postulated
that the laws of geometric dynamics in special relativity are invariant
under the Poincar{é} group $\mathscr{P}$. However, at the level
of high energy physics, it was recognized  that the invariance
of laws under the full isometry group $\mathscr{P}$ can not be taken
as granted. As early as in 1949, Dirac \cite{Dirac1949} wrote that
``I do not believe there is any need for physical laws to be
invariant under these reflections\footnote{The space and time inversions.}, although all the exact laws of
nature so far known do have this invariance.'' In 1956 Lee and Yang
\cite{Lee-Yang1957} suggested explicitly that parity conservation
might be violated in weak interactions by analyzing the data which
demonstrated meson states $\tau$ and $\theta$ with almost identical
mass but decayed to final states of opposite parity, which was shortly
verified experimentally by Wu \cite{Wu-}. On the other hand, all
known experimental data in high energy physics support the Lorentz
invariance principle (in the context of special relativity): laws
of Nature are invariant under the sub-group $\mathscr{L}_{0}$ of
all proper and orthochronous Lorentz transformations and the translations.

The strong form of the Lorentz invariance perhaps is good for dynamics,
however it is violated at the sub-atomic level, the principle of Lorentz
invariance, which should be applicable for high energy physics with
the general theory of relativity, has not been examined in literature.
In the past, according to Dirac \cite{Dirac 1958} for example, there
was no need to consider gravitational effect at the sub-atomic level, and
therefore the principle of Lorentz invariance was not in demand.
However, the successful detection of gravitational wave radiations
originated from a pair of merging black holes (see \cite{castelvecchi,LIGO})
and the more recent detection (see \cite{Science1,Science2,Science3,Science4,Science5})
from a neutron star merger confirmed one of predictions made by Einstein
\cite{Einstein-wave} derived from the general theory of relativity
\cite{Einstein-general,Einstein-meaning}, and understanding the mechanics
of gravitational waves requires high energy physics and the theory
of gravitation, in which the Lorentz invariance plays an important
role. The aim of this note is to formulate a version of the principle
of Lorentz invariance within the general theory of relativity. A correct
formulation of the Lorentz invariance is crucial for determining
laws of Nature.

\section{Spacetime orientation as causal structure}

Let us now consider a spacetime $(M,g_{\mu\nu})$. In order to develop
a theory of Lorentz invariance with the general relativity, a careful
study of causal structure of spacetime seems necessary. In history,
the \emph{causality} as an important component of the spacetime model
emerged, surprisingly, firstly not from the investigation of cosmology
and gravitation, but from the high energy physics. The current knowledge
supports the postulate that it is the causality structure of the spacetime
which is responsible for the concepts of anti-matters and violations
of various discrete symmetrices. The causality structure became prominent,
largely due to the discovery of singularities of the spacetime in
1960's in the seminal work by Hawking and Penrose (see e.g. \cite{Penrose1965,Hawking1967,Hawking-Penrose1970,Hawking-Ellis}).
Here we only need one element of causal structures which is related
to spacetime orientation.

The best way to describe spacetime orientation is to use the language
of principal fiber bundles (see e.g. Hirzebruch \cite{Hirzebruch}).
Let $P(M)$ denote the fiber bundle of all frames (also called tetrads)
$(x,(e_{k}))$, where $(e_{k})$ is a linear basis of the tangent
space $T_{x}(M)$ at point $x\in M$. Suppose $(x^{\mu})$ is a coordinate
system near $x$, then a linear basis can be expressed as $e_{k}=x_{\;k}^{l}\frac{\partial}{\partial x^{l}}$
in terms of partial coordinate derivatives, and therefore, it leads
to a coordinate system $(x^{\mu},x_{\;k}^{l})$ for the frame bundle.
$P(M)$ is a principle fiber bundle with its structure group the general
linear group $\textrm{GL}(4)$, so that, if $a=(a_{\;j}^{i})$ is an
invertible real $4\times4$ matrix, then its (effective) right action
sends a frame $(e_{k})$ to $(a_{\;k}^{j}e_{j})$, which leads to
the transformation of the corresponding coordinates 
\[
(x^{\mu},x_{\;k}^{l})\longrightarrow\left(x^{\mu},a_{\;k}^{j}x_{\;j}^{l}\right).
\]
In this sense, $P(M)$ is a principal fiber bundle over $M$ with
its structure group and canonical fiber $\textrm{GL}(n)$.

The Lorentz metric $(g_{\mu\nu})$ determines a fiber sub-bundle $O(M)$
consisting of all orthonormal frames $(x,(e_{k}))$ where $g(e_{\sigma},e_{\rho})=\eta_{\sigma\rho}$.
$O(M)$ possesses a differential structure as the sub-manifold of
dimension 10, defined by the following ten (independent) equations:
\begin{equation}
g_{\mu\nu}(x)x_{\;\sigma}^{\mu}x_{\;\rho}^{\nu}=\eta_{\sigma\rho}.\label{eq:o-bundle}
\end{equation}
It is easy to verify that the right action under $(a_{\;k}^{j})$
leaves $O(M)$ invariant if and only if $a_{\;\mu}^{\sigma}\eta_{\sigma\rho}a_{\;\nu}^{\rho}=\eta_{\mu\nu}$,
that is $a$ is a Lorentz matrix. Therefore $O(M)$ is a principal
fiber bundle with its structure group and typical fiber the Lorentz
group $\mathscr{L}$.

If $(x,(e_{\alpha}))$ and $(x,(\tilde{e}_{\alpha}))$
are two orthonormal frames of $T_{x}M$, then $\tilde{e}_{\mu}=\varLambda_{\;\mu}^{\nu}e_{\nu}$
where $\varLambda$ is a Lorentz transformation. It is said two frames
$(x,(e_{\alpha}))$ and $(x,(\tilde{e}_{\alpha}))$ are equivalent
if $\tilde{e}_{\mu}=\varLambda_{\;\mu}^{\nu}e_{\nu}$ with $\varLambda\in\mathscr{L}_{0}$.
Thus, for every $x\in M$, the fiber $O(M)_{x}$ is decomposed into
a direct sum of 4 disjoint connected components: $O(M)_{x}=\left[e\right]\cup\left[\mathtt{P}e\right]\cup\left[\mathtt{T}e\right]\cup\left[\mathtt{PT}e\right]$
where $e\in O(M)_{x}$ (where $\mathtt{P}$ and $\mathtt{T}$ denote
the space and time inversions respectively), which in turn leads to
two principal bundles. Let $O(M)/\mathscr{L}_{0}$ denote the set
of equivalence classes, and the natural projection is denoted by $\sigma:O(M)\rightarrow O(M)/\mathscr{L}_{0}$
sending every element of $O(M)$ to its equivalent class. Then $O(M)$
is a principal fiber bundle over $O(M)/\mathscr{L}_{0}$. On the other
hand, the natural projection $\pi$ from $O(M)$ to $M$ defines a
natural projection from $O(M)/\mathscr{L}_{0}$ to $M$, and $O(M)/\mathscr{L}_{0}$
is a principal fiber bundle over $M$ with its structure group $\mathscr{L}$
and typical fiber $\mathscr{L}/\mathscr{L}_{0}$ which is a discrete
group of four elements.

It is said that a spacetime orientation exists on the spacetime $M$,
if the structure group $\mathscr{L}$ of $O(M)$ can be reduced to
the Lorentz subgroup $\mathscr{L}_{0}$, in the sense that there is
a principal subbundle $L(M)$ of $O(M)$ over $M$ with both its
structure group and typical fiber being $\mathscr{L}_{0}$. According
to Hirzebruch (\cite{Hirzebruch} Theorem 3.4.5 on page 45), the structure
group $\mathscr{L}$ of $O(M)$ can be reduced to $\mathscr{L}_{0}$,
if and only if there is a global (smooth) section of the principal
fiber bundle $O(M)/\mathscr{L}_{0}$ over $M$. A (smooth) section
of the principal fiber bundle of $O(M)/\mathscr{L}_{0}$ is called
a spacetime orientation of the spacetime $(M,g_{\mu\nu})$.

\section{Principle of Lorentz invariance}

To be able to implement the principle of Lorentz invariance in the
general theory of relativity, we work with a model of spacetime, a
four dimensional manifold $M$ endowed with a Lorentz metric $(g_{\mu\nu})$,
such that the structure group $\mathscr{L}$ of the orthonormal frame
bundle $O(M)$ can be reduced to the Lorentz subgroup $\mathscr{L}_{0}$,
that is, there is a global smooth section of the bundle $O(M)/\mathscr{L}_{0}$.
Suppose $s:M\rightarrow O(M)/\mathscr{L}_{0}$ is such a section which
determines a spacetime orientation of $M$. Suppose $F:M\rightarrow M$
is a diffeomorphism preserving the gravitational field $(g_{\mu\nu})$,
so that the tangent mapping $F_{\star}$ sends the fiber $O(M)_{x}$
to the fiber $O(M)_{F(x)}$for every $x\in M$. Let $s(x)=\left[e(x)\right]$
for every $x\in M$, represented by the equivalent class of a frame
$(x,(e_{\mu}(x)))$, and let $\tilde{e}(x)=F_{\star}(e(x))$ which
belongs to $O(M)_{F(x)}$, where $\tilde{e}_{\mu}(x)=F_{\star}\left(e_{\mu}(x)\right)$
for every $x\in M$. We say an isometry $F$ is proper and orthochronous
(with respect to the spacetime orientation $s$), if $\left[\tilde{e}(x)\right]=\left[e(F(x))\right]$
for every $x$, that is, the spacetime orientation is invariant under
the differential map $F_{\star}$.

We are now in a position to formulate the principle of Lorentz invariance:
laws of Nature are invariant under any proper and orthochronous
isometry of $(M,g_{\mu\nu})$. That is, laws of Nature take the same
mathematical formulation under any proper and orthochronous isometry
of $(M,g_{\mu\nu})$.

We conclude this note by raising the following question, which should
be worthy of study. Recall that Einstein's field equation and the
geometric dynamics were derived under the general version of Lorentz
invariance, which is violated at the sub-atomic level. Of course, there
is no reason, though there is no any experimental evidence yet, to
explain why the general Lorentz invariance is not violated at the
level of geometric dynamics too. This consideration leads to the following
question, is there any need to modify the geometric dynamics in particular
Einstein's field equation, if only the principle of Lorentz invariance
as formulated above is accepted?

\small

{\small{}\vskip0.5cm }\textit{\small{}{Acknowledgements}}{\small{}.
The author would like to thank Dominic Joyce for the helpful comments.}{\small \par}

\end{document}